\documentclass[]{aa} 
\usepackage{graphicx}
\usepackage{txfonts}

\begin{document}
   \title{Testing the comet nature of main belt comets. The spectra of 133P/Elst-Pizarro and 176P/LINEAR.}
   \author{J. Licandro
	  \inst{1,2}
	  \and
          H. Campins
          \inst{3}
          \and
          G. P. Tozzi
          \inst{4}
          \and
          J. de Le\'on
          \inst{5}
          \and
          N. Pinilla-Alonso
          \inst{6}
	    \and
	    H. Boehnhardt
          \inst{7}
	  \and
	  O.R. Hainaut
	          \inst{8}
 }

   \offprints{J. Licandro}

  \institute{Instituto de Astrof\'{\i}sica de Canarias, c/V\'{\i}a L\'actea s/n, 38200 La Laguna, Tenerife, Spain. \\
              \email{jlicandr@iac.es}
              \and 
              Departamento de Astrof\'{\i}sica, Universidad de La Laguna, E-38205 La Laguna, Tenerife, Spain\\
              \and
              Physics Department, University of Central Florida, Orlando, FL, 32816, USA.\\
              \and
              INAF -- Osservatorio Astrofisico di Arcetri, I-50125 Firenze, Italy\\
              \and
              Instituto de Astrof\'{\i}sica de Andaluc\'{\i}a, Granada, Spain.\\
              \and
              NASA Postdoctoral Program at NASA Ames Research Center, Moffett Field, CA, USA\\
              \and
              MPS, Katlenburg-Lindau\\
              \and
              ESO, Karl Schwarzschild Stra\ss e , 85748 Garching bei M\"unchen, Germany.
 }
   \date{Received August 2010; accepted}

 
  \abstract
   {Dynamically, 133P/Elst-Pizarro and 176P/LINEAR are main belt asteroids, likely members of  the Themis collisional family, and unlikely of cometary origin. They have been observed with cometary-like tails,  which may be produced by water-ice sublimation. They are part of a small group of objects called Main Belt Comets (MBCs, Hsieh \& Jewitt 2006). }
   {We attempt to determine if these MBCs have spectral properties compatible with those of comet nuclei or with other Themis family asteroids.}
   {We present the visible spectrum of MBCs 133P and 176P, as well as  three Themis family asteroids: (62) Erato, (379) Huenna and (383) Janina, obtained in 2007 using three telescopes at ``El Roque de los Muchachos"' Observatory, in La Palma, Spain , and the 8m Kueyen (UT2) VLT telescope at Cerro Paranal, Chile. The spectra of the MBCs are compared with those of the Themis family asteroids, comets,  likely "dormant" comets and asteroids with past cometary-like activity in the near-Earth (NEA) population. As 133P was observed active, we also look for the prominent CN emission around 0.38 $\mu$m typically observed in comets, to test if the activity is produced by the sublimation of volatiles.}
   {The spectra of 133P and 176P resemble best those of B-type asteroid and  are very similar to those of Themis family members and another activated asteroid in the near-Earth asteroid population, (3200) Phaethon.  On the other hand, these spectra are significantly different from the spectrum of comet 162P/Siding-Spring and most of the observed cometary nuclei. CN gas emission is not detected in the spectrum of 133P. We determine an upper limit for the CN production rate Q(CN)  $= 2.8 \times 10^{21}$ mol/s, three orders of magnitude lower than the Q(CN) of Jupiter family comets observed at similar heliocentric distances. }
   {The spectra of 133P/Elst-Pizarro and 176P/LINEAR confirm that they are likely members of the Themis family of asteroids, fragments that probably retained volatiles, and unlikely have a cometary origin in the trans-neptunian belt or the Oort cloud.
They have similar surface properties to activated asteroids in the NEA population, which supports the hypothesis that these NEAs are scattered MBCs.
The low Q(CN) of 133P means that, if water-ice sublimation is the activation mechanism, the gas production rate is very low and/or the parent molecules of CN present in the nuclei of normal comets are much less abundant in this MBC.
}

   \titlerunning{The spectra of 133P/Elst-Pizarro and 176P/LINEAR. }
   \maketitle
   

\section{Introduction}

Comets are known to originate in the trans-neptunian belt (TNB) and the Oort Cloud. Observationally, comets  are distinguished from asteroids by the presence of a coma and/or tail.  However, this is not a conclusive criterion. e.g. icy objects only develop a coma if the temperature is sufficient to sublimate ices, so distant comets may not show activity.

For the purpose of this paper we define that an object has a cometary origin if it shows similar physical properties to objects that have been considered comets up to now, i.e., icy objects scattered from the trans-neptunian belt or the Oort cloud. 

The Tisserand parameter with respect to Jupiter ($T_J$) provides a simple way to discriminate dynamically between asteroids and comets (Kresak \cite{Kresak82}; Kosai \cite{Kosai92}). Main-belt asteroids move in orbits with $T_J$ $>$ 3, while comets have unstable orbits with $T_J$ $<$ 3.

Recently, some objects in orbits indistinguishable  from those of other main belt asteroids have been observed ``active" (with a dust coma and/or tail). They have been called Main Belt Comets (MBCs) (Hsieh \& Jewitt \cite{Hsieh06}).  133P/Elst-Pizarro (hereafter, 133P) and 176P/LINEAR (hereafter, 176P) are also identified as asteroids 7968 and 118401 respectively. They have $T_J$ = 3.184 and 3.166 respectively and they are two of the seven MBCs observed so far . 

Dynamical simulations show that MBCs are extremely unlikely to originate in the TNB or the Oort Cloud (e.g., Fern\'andez et al. 2002), indicating that they likely formed in situ. Recent work suggests that some icy trans-neptunian objects (TNOs) might have been delivered to the asteroid belt during the Late Heavy Bombardment (Levison et al. \cite{levison2009}), but even those simulations fail to produce low inclination, low-eccentricity orbits such as those of 133P and 176P. 

133P is the first discovered MBC (Elst et al. \cite{Elst96}) and, so far, the best characterized one. Since then, it has been observed active every perihelion passage around the orbital quadrant following perihelion, and inactive the two quadrants around aphelion (Boehnardt et al. \cite {Boehnhardt1997}; Toth \cite{Toth06}, Hsieh et al. \cite{Hsieh04, Hsieh09, Hsieh10}, Bagnulo et al. \cite{Bagnulo2010}). This recurrent activity supports the hypothesis that 133P has an ice reservoir, which periodically sublimates and elevates dusty material into the coma and tail region. The dust coma and tail, and not a gas emission feature, are so far the only signatures of ongoing activity in MBCs
On the other hand, 176P is the third known MBC and it exhibited cometary activity only at the time of its discovery in 2005 (Hsieh et al. \cite{Hsieh06}). It was never reported active since.

Understanding the origin of these bodies is crucial. If they are formed ``in situ'' and, in particular, if they are members of a collisional family and if their activity is due to water ice sublimation, there should be water ice in many asteroids. If they are captured TNB or Oort cloud comets, the mechanisms that drove them to their present orbits needs to be understood.
The existence of ice in main-belt objects is surprising given their proximity to the Sun, and presents intriguing opportunities for constraining the temperature, compositional and structure of primitive asteroids and our protoplanetary disk. Recent theoretical model calculations indicate that subsurface ice can survive in an MBC over the lifetime of the solar system (Schorghofer \cite{Schorghofer}).

 The R-band albedo of 133P and 176P is $p_R=0.05\pm0.02$ and $p_R=0.06\pm0.02$ respectively (Hsieh et al. \cite{Hsieh09}). The 133P albedo determination is in agreement with more recent estimates by Bagnulo et al. (\cite{Bagnulo2010}),  using the polarization vs phase function and assuming that the relation with the albedo is the same of that found for asteroids ($p_R=0.07\pm0.01$). All these values are consistent with the albedos of primitive  C-, B-, D-type asteroids and cometary nuclei (Fern\'andez et al. \cite{Fernandez05}), implying that the albedo does not appear to be a decisive diagnostic parameter for determining the source zone of MBCs. 

The orbits of 133P and 176P are typical of Themis collision family members, or a sub-family of it, called the Beagle family (Nesvorny et al. \cite{Nesvorny2007}). The Themis family is likely the result of the breakup of a parent asteroid about 400 km in diameter $\sim1$ Gyr ago (Marzari et al. \cite{Marzari1995}; Tanga et al. \cite{Tanga1999}; Nesvorny et al. \cite{Nesvorny2007}) and is one of the largest and statistically most robust asteroid families (e.g., Carusi \& Valsecchi \cite{Carusi1982}; Zappal\`a et al. \cite{Zappala1990}). The surface of 133P and 176P is neutral in color (Hammergren 1996; Hsieh et al. \cite{Hsieh10}; Bagnulo et al. \cite{Bagnulo2010}), consistent with  the nearly neutral reflection spectra of primitive asteroids of the C-complex, but bluer than the colors of the majority of cometary nuclei (Lamy and Toth \cite{Lamy2009}). 
Given their origin from a common parent, Themis family members are thought to be compositionally homogeneous, as corroborated by spectroscopical studies showing that the family is dominated by primitive C-type asteroids that also exhibit signs of aqueous alteration (Bell \cite{Bell1989}; Florczak et al. \cite{Florczak1999}; Ivezic et al. \cite{Ivezic2002}; Moth\'e-Diniz et al. \cite{Moth2005}). The detection of water ice on the surface of  the largest member of the family (24) Themis (Campins et al. \cite{campins2010}, Rivkin \& Emery et al. \cite{rivkin2010}) and the evidence of aqueous alteration in about 50\% of the Themis family asteroids strongly support that water is widely present in the Themis family. 

In this paper we present and study spectra of two MBCs, 133P and 176P to address the open question of the origin of this population and, in particular, to discriminate if they are Themis family asteroids or interloper comets . In section 2 we present visible spectra of both MBCs obtained using different telescopes,  aditionaly we present new spectra of three Themis family asteroids. In section 3 we compare the spectra of the MBCs with spectra of objects with asteroidal and cometary nature: the spectra of Themis family asteroids, the spectrum of (3200) Phaethon, a prototype of near-Earth asteroid that presented cometary-like activity in the past and possibly linked to MBCs (Meng et al. \cite{Mengetal04}; Licandro et al. \cite{Licandroetal07}); and the spectra of comet nuclei. In Sect. 4 we determine an upper limit for the CN production rate of 133P, and we present our conclusions in Sect. 5.

\section{Observations and data reduction}

Visible spectra of 133P were obtained during its 2007 perihelion passage using three telescopes: the 3.5m Telescopio Nazionale Galileo (TNG) and the 4.2m William Herschel Telescope (WHT) at ``El Roque de los Muchachos" Observatory (ORM, Canary Islands, Spain), and the 8m Kueyen (UT2) VLT telescope at Cerro Paranal (Chile). On the other hand, visible spectra of 176P were obtained on January 20, 2007 using the 2.5m Nordic Optical Telescope (NOT) at the ORM.  

The DOLORES spectrograph at the TNG was used on June 17, 2007 to observe 133P. Spectra using the LR-B and the LR-R grisms were obtained, with dispersions of 2.52 and 2.61 $\AA/$pixel respectively. The 5.0"  slit width was used, oriented at the parallactic angle to minimize the slit losses due to atmospheric dispersion. Three spectra with an exposure time of 600s each of 133P where obtained with the LR-R grism, covering the  0.52 -- 0.92$\mu$m spectral range. The object was shifted in the slit  direction by 5'' between consecutive spectra to better correct the fringing. Two spectra of 1200s each of 133P where obtained with the LR-B grism, covering the  0.37 to 0.70$\mu$m spectral range.

A spectrum of 133P was also obtained with the WHT on Jun 17, 2007, using the double arm ISIS spectrograph. Spectra in the red and blue arms where obtained simultaneously, using the R300B grating in the blue arm, with a dispersion of 0.86$\AA/$pixel, and the R316R grating in the red arm, with a dispersion of 0.93$\AA/$pixel. A 3" slit width was used, oriented at parallactic angle. A single 900s exposure was done in both arms. 133P was observed again on July 9, 2007 with the WHT using ISIS. Spectra in the red and blue arms where obtained simultaneously, using the R300B grating in the blue arm, and the R158R grating in the red arm (with a dispersion of 1.81$\AA/$pixel). A 5" slit width was used oriented at the parallactic angle. Two spectra of 1800s where obtained in both arms. 

Finally, 133P was observed on July 21, 2007 with the VLT using the camera spectrograph FORS2. A total of 7 low-resolution spectra of 300s each were obtained. The grism 150I without an order sorting filter and a slit of 1.3" width were used. The covered spectral range was 0.33 -- 0.66 $\mu$m with a dispersion of 230 \AA/mm, corresponding to about 3.5 \AA/pixel. Solar analogue star HD209847 was also observed in order to obtain the reflectance spectrum.
The observations were done in service mode, and all the spectra were recorded at airmass  $>1.6$. Since the slit was not oriented in the parallactic angle, the spectra suffered from strong atmospheric refraction and they could not be used to measure the spectral reflectance. But the signal-to-noise ratio (SNR) of the obtained spectrum was sufficiently hight to search for the presence of gas emission, as that of CN at 3880 \AA ~(see Sect. 4). To do that, the spectrum of spectrophotometric standard LTT 377  was obtained for calibration purposes using the same configuration. 



On January 20.22 UT, 2007, 176P was observed with the 2.5m NOT telescope using the camera-spectrograph ALFOSC. The 5.0" slit width was used, oriented in the parallactic angle. Two spectra of 1800s were obtained with the \#4 grism, covering the  0.32 -- 0.91$\mu$m spectral range, with a dispersion of 3 \AA/pixel.


Images were overscan and bias corrected, and flat-field corrected using lamp flats. The two-dimensional spectra were extracted, sky background subtracted, and collapsed to one dimension. The wavelength calibration was done using the Neon and Argon lamps.  The spectra of each asteroid, obtained at different positions in the slit, were averaged. 

\begin{figure}
	\centering
	\includegraphics[width=6cm, angle=0]{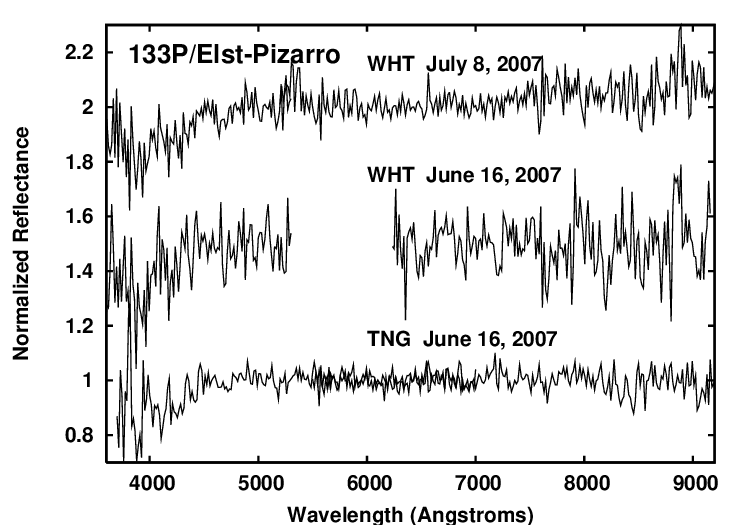}
	\caption{Spectra of 133P/Elst-Pizarro obtained with the TNG and WHT telescopes. Spectra are normalized to unity at 5500 \AA~ and shifted vertically by 0.3 for the sake of clarity. The spectrum obtained on June 16 with the WHT (middle one) does not cover the whole spectral range (there is a "hole" between 5400 and 6200 \AA), in this case both parts (red and blue arm) are scaled using TNG spectrum obtained in the same night. }
	\label{Fig1}
\end{figure}

\begin{figure}
	\centering
	\includegraphics[width=6cm, angle=0]{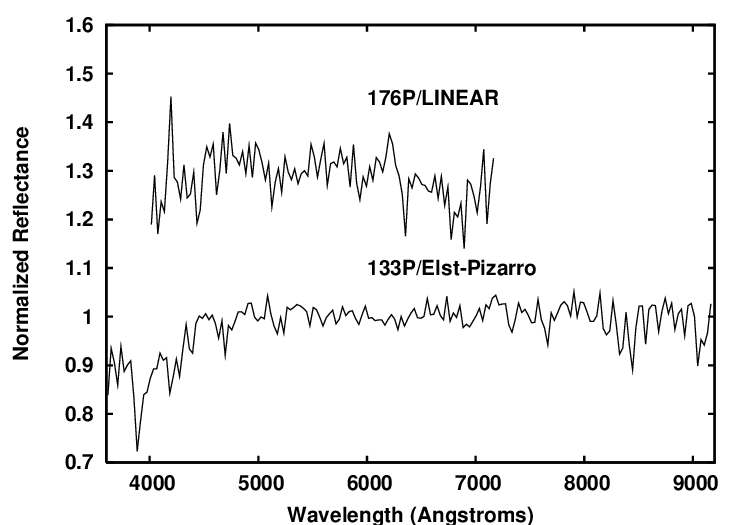}
	\caption{Spectrum of 176P/LINEAR obtained with the NOT telescope binned to a resolution of 30\AA, compared with the mean spectrum of  133P/Elst-Pizarro also binned to the same resolution. Spectra are normalized to unity at 5500 \AA~  and shifted vertically by 0.3 for clarity. The spectrum of 176P is plotted only in the 4000-7200 \AA~ range as the SNR outside this spectral region is very low. Notice that the spectra of 133P and 176P are similar, both show the slightly and rare negative (blue) slope in this wavelength region, typical of B-type asteroid. }
	\label{Fig2}
\end{figure}

To correct for telluric absorption and to obtain the relative reflectance, a G star from the Landolt list (Landolt, \cite{landolt}) was observed at different airmases (similar to those of the object)  before and after the asteroid's observations, and used as a solar analogue star. Landolt (SA) 115-271, and Landolt (SA) 107-684 were observed to correct the spectra of 133P and 176P, respectively. The spectrum of each solar system object was divided by that of the corresponding solar analogue star, and then normalized to unity around 0.55$\mu$m, thus obtaining the normalized reflectance. The obtained spectra of 133P are shown in Fig. \ref{Fig1}. The spectrum of 176P is shown in Fig. \ref{Fig2} together with the weighted mean spectrum of the 3 spectra of 133P shown in Fig. \ref{Fig1}. In Fig. \ref{Fig2} both spectra are rebinned to a resolution of 30\AA. The spectrum of 176P is plotted only in the 0.4--0.72 $\mu$m region because of the very low SNR at bluer wavelengths and the strong fringing in the red.

\begin{figure}
	\centering
	\includegraphics[width=6cm, angle=0]{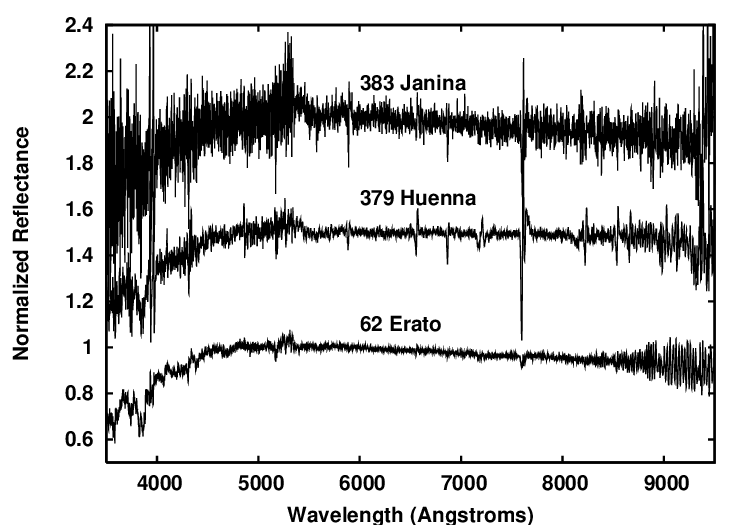}
	\caption{Spectra of 3 Themis family asteroids obtained with the WHT. Spectra are normalized to unity at 5500 \AA~ and shifted vertically by 0.5. Notice they are all very similar.}
	\label{Fig3}
\end{figure}

Also asteroids (62) Erato, (379) Huenna and (383) Janina, all of them considered to be members of the Themis family, were observed  on April 1, 2006 with the  WHT using the ISIS spectrograph. Spectra in the red and blue arms were obtained simultaneously, using the R300B and R158R gratings respectively. A 5" slit was used oriented in the parallactic angle.  Solar analogue star  BS 4486 and G-type Landolt star SA 107-689  were observed to obtain the relative reflectance of the asteroids. The resulting reflectance spectra, processed in a similar way as those of 133P and 176P,  are shown in Fig.\,\ref{Fig3}. The spectra of these Themis family asteroids is used to compare with that of 133P and 176P.



\section{Data analysis}

The spectra of 133P and 176P are very similar,  presenting a neutral to slightly bluish slope at wavelengths larger than 5000\AA, and a drop in the UV region. This behavior is typical of B-type asteroids, according to the Tholen spectrophotometric classification system (Tholen \cite{Tholen1989}). Measuring the spectral slope in the visible region ($S'_V$) as in Licandro et al. (\cite{licandro08}), we obtain $S'_V$ = 0.00 $\pm 0.01$~\%/1000\AA ~for 133P and $S'_V$ = -0.03 $\pm 0.01$~\%/1000\AA~ for 176P. 176P is spectroscopically a B-type asteroid, while the neutral slope of 133P place it on the line between a B- or a C-type. B-, C- and F-type asteroids belong to the C-complex, {which comprises} asteroids with a primitive composition and very low albedo, usually linked to the CI, CM carbonaceous chondrites. 

Bagnulo et al. (\cite{Bagnulo2010}), based on photometric and polarimetric measurements concluded that the overall light scattering behavior (photometry and polarimetry) of 133P is different from that of cometary nuclei  and resembles most closely that of F-type asteroids. However, the most distinctive spectral feature of the F-type asteroids, as compared to other types, is the absence of an UV absorption (Tholen \cite{Tholen1989}) like that observed in the spectra of  133P and 176P. 

Recently, Clark et al. (\cite{Clark2010}) and de Le\'on et al. (\cite{deLeon2011}) analyzed visible and near infrared spectra (covering the 0.5 to 2.5 $\mu$m region) of asteroids classified as B-type  from their visible spectra.  Both papers point out  that the spectra of B-type asteroids show a range of behaviors in the infrared region. Hence, B-type asteroids can have different light scattering properties and some of these asteroids, as could be the case of 133P (Bagnulo et al. \cite{Bagnulo2010}), can have photometric and polarimetric properties that resemble most closely those of F-type asteroids. 

\subsection{MBCs and the Themis family}


As mentioned in the introduction, asteroids 133P and 176P can show cometary-like tails. According to their dynamical properties, they also belong to the Themis family (Haghighipour, \cite{Haghighipour}). Therefore, we first compare their spectra with the spectra of other Themis family asteroids to see if these MBCs are compatible with being fragments produced by the collision that produced the family.
  
\begin{figure}
	\centering
	\includegraphics[width=6cm, angle=0]{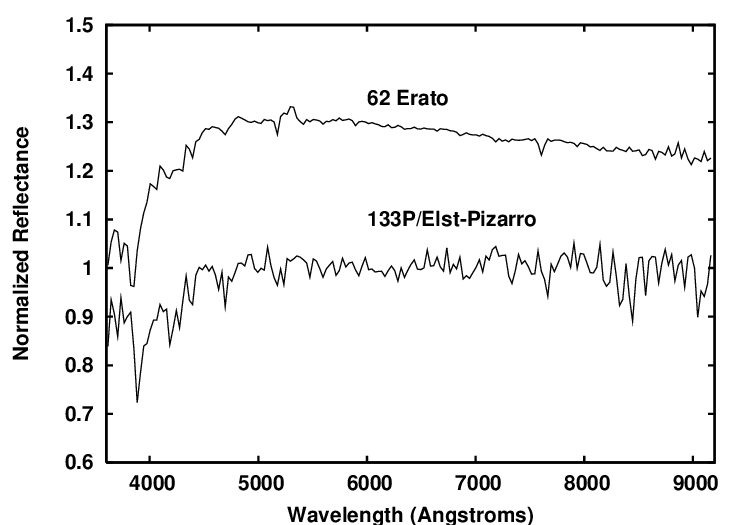}
	\caption{ Visible spectra (62) Erato and 133P/Elst-Pizarro. The spectra are normalized to unity at 5500 \AA~ and shifted vertically for clarity.}
	\label{Fig4}
\end{figure}

The Themis family is compositionally primitive, and the taxonomy of its family members is very homogeneous. Themis family asteroids are mainly C- and B-types in the Tholen classification system (e.g., Mothe-Diniz et al. \cite{Moth2005}, Florczak et al. \cite{Florczak1999}). Considering Bus taxonomy (Bus \cite{Bus99}), there are 36 asteroids  belonging to the C complex (6 C-type, 17 B-type, 5Ch-type, and 8 Cb-type) and 7 asteroids to the X complex (5 X-type, 1 Xc-type, and 1 Xk-type). This is compatible with the breakup of a large ($\sim$ 370 km according to Tanga et al. \cite{Tanga1999}) C-class parent body.
The spectra of the three Themis family asteroids presented here (62 Erato, 379 Huenna and 383 Janina) are very similar, typical of B-type asteroids. In Fig.\,\ref{Fig4} we compare the spectrum of 133P with that of (62) Erato, which is the one with the best SNR. Notice that  the spectra of 133P and (62) Erato are very similar within the uncertainties, supporting that 133P and 176P are likely fragments of the Themis family parent body. The spectrum of (62) Erato is slightly bluer ($S'_V=0.02\pm0.01$~\%/1000\AA) than that of 133P ($S'_V=0.00\pm0.01$~\%/1000\AA), but similar within the errors, and both have a pronounced absorption feature in the UV.

In addition, Rousselot et al. (\cite{Rousselotetal11}) present a near-infrared spectrum of 133P in the 1.1-2.4 $\mu$m region that it is compatible with those of Themis-like B-types in Clark et al. (\cite{Clark2010}).

\subsection{MBCs and activated asteroids in the Near-Eart objects (NEO) population.}

MBCs are not the only objects in asteroidal orbits that present cometary-like activity. There is a group of asteroids in the Near-Earth population with associated meteor showers: (3200) Phaethon, 2001 YB$_5$ and 2005 UD (Whipple \cite{Whiple93}, Meng et al. \cite{Mengetal04} and Ohtsuka et al. \cite{Ohtskuaetal06}). Meteor showers are usually associated to comets, and in the case of these asteroids, it is indicative of a past cometary-like activity. The relation between (3200) Phaethon and its associated meteor shower, the Geminds, has been well studied by Gustafson (\cite{Gusta89}) and Williams \& Wu (\cite{WillWu93}). 

There is also one asteroid, (4015) Wilson-Harrington (also known as comet 107P) that was first discovered as an active comet, then re-discovered as an asteroid (Bowell et al. \cite{Bowell92}), and that has never been observed active again. According to Campins \& Swindle (\cite{CampSwin98}), 107P is a potential meteorite producing object. 


Meng et al. (\cite{Mengetal04}) noticed that all these ``activated asteroids" in the NEA population have spectra and colors compatible with B-type asteroids. Licandro et al. (\cite{Licandroetal07}) suggested a link between activated asteroids in the NEA population and MBCs, as their colors and/or spectra correspond to those of Cb- or B-type asteroids, therefore, the activated asteroids in the NEA population are likely MBCs scattered from the MB to NEA orbits. 

\begin{figure}
	\centering
	\includegraphics[width=6cm, angle=0]{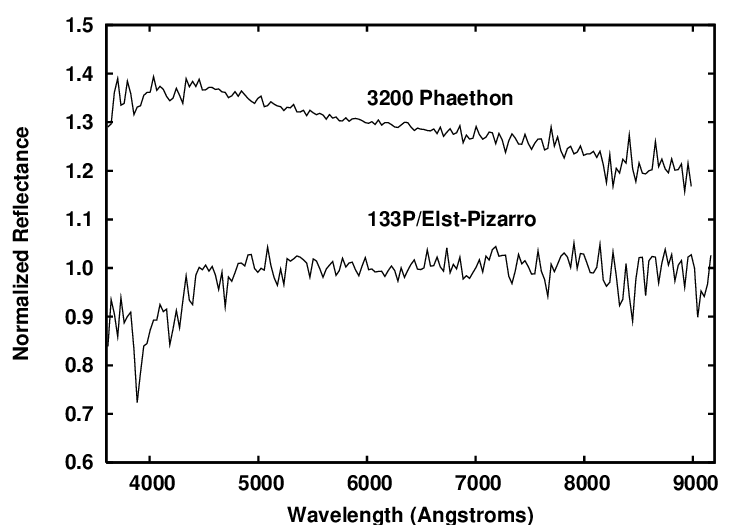}
	\caption{ Visible spectra of near-Earth asteroid (3200) Phaethon and 133P/Elst-Pizarro. The spectra are normalized to unity at 5500 \AA~ and shifted vertically for clarity.}
	\label{Fig5}
\end{figure}

In  Fig.\,\ref{Fig5} the spectra of 133P and (3200) Phaethon (from Licandro et al. \cite{Licandroetal07}) are plotted together. The spectrum of 133P has similarities with the spectrum of (3200) Phaethon: both present the slightly blue slope in the 5000-9000~\AA~ spectral region and a drop of in the UV below 4500~\AA. Anyway, there are some differences indicative of possible differences in the composition of the surface of both objects: (3200) Phaethon spectrum is slightly bluer ($S'_V$=0.04~\%/1000\AA) than 133P one, but with a slope similar to the spectrum of 176P ($S'_V$=0.03~\%/1000\AA); the UV absorption in the spectrum of (3200) Phaethon is not as well defined as in the spectrum of 133P. 

Licandro et al. (\cite{Licandroetal07}) argued that (3200) Phaethon's surface has abundant phyllosilicates and that this object is unlikely to have a cometary origin. The similarity between (3200) Phaethon's spectrum and that of 133P and 176P then also supports an asteroidal nature for these MBCs. 


Recent results show that (3200) Phaethon is likely to be a fragment of asteroid (2) Pallas family (de Le\'on et al. \cite{leon2010}), in fact, its visible spectrum is identical to those of the Pallas family asteroids. The activity of (3200) Phaethon, 133P and 176P suggest that both, (2) Pallas and (34) Themis, produced fragments with the ability of develop cometary-like activity.

\subsection{MBCs and comets.}

\begin{figure}
	\centering
	\includegraphics[width=6cm, angle=0]{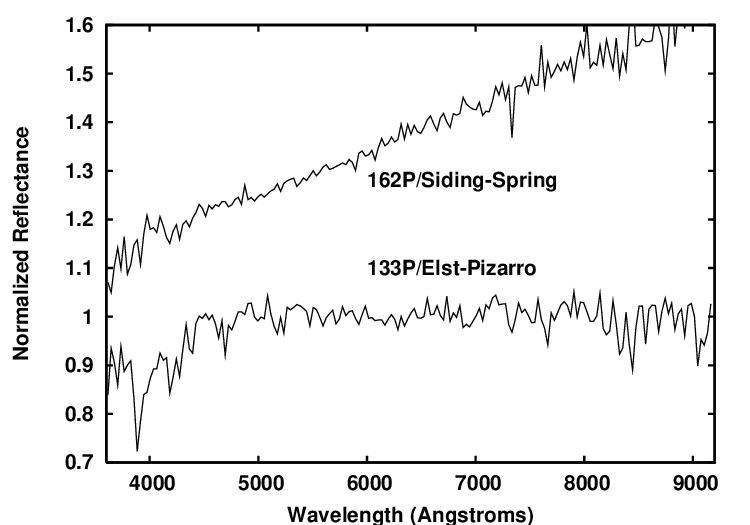}
	\caption{Visible spectra of comet nucleus 162P/Siding-Spring and 133P/Elst-Pizarro. Spectra are normalized to unity at 5500 \AA~ and shifted vertically for clarity.}
	\label{Fig6}
\end{figure}

To explore the possible cometary origin of 133P, we plot in Fig.\,\ref{Fig6} the spectrum of 162P/Siding-Spring from Campins et al. (\cite{campins2006}), one of the best SNR spectrum of a comet nucleus, rebinned to the same spectral resolution as the other objects presented in this paper. Notice that 162P's spectrum is very different to that of the Themis family asteroids and (3200) Phaethon. The spectrum of 162P is compatible with the spectra of  D-type asteroids: featureless with a steep red slope ($> 0.07$~\%/1000\AA) and with no absorption feature in the UV. The large majority of comet nuclei with observed spectra have a spectrum similar to that of 162P and compatible with that of P- or D-type asteroids (Jewitt \cite{jewitt2002},  Licandro et al. \cite{licandro2002}, Campins et al. \cite{campins2006}, Campins et al. \cite{campins2007}
Snodgrass et al. \cite{Snodgrass08}). Also the asteroids in cometary orbits (objects with an asteroidal aspect but with cometary-like orbits) that more likely have a cometary origin (those with T$_J<$2.7) are all P- and D-types (Licandro et al. \cite{licandro08}).  P- and D-type  asteroids have featureless spectra, with a slightly red to very red slope ($S'_V >$ 0.02~\%/1000\AA), and it is assumed that they are composed of primitive material. The fact that the MBCs and comets present spectra of different taxonomical types argues also against a cometary origin of the MBCs.

\section{An upper limit for the CN production rate of 133P/Elst-Pizarro}


Even if, as mentioned in Sect. 2, the spectra of 133P obtained with the VLT are not useful in getting the reflectance, they can be used to look for any signature of gas emission typically observed in cometary comae, that of CN at $\sim$3800\AA ~being the most prominent one. When observed with the VLT, 133P presented a faint tail and a diffuse dust coma, so the detection of any gas emission would be  a strong test to determine if the observed tail and coma were produced by ice sublimation as in normal comets.

The spectra of 133P obtained with the VLT were calibrated in intensity (erg/s cm$^2$ \AA\ ster) and in projected distance (km) along the slit, using the spectrum of the spectrophotometric standard. To increase the signal to noise ratio all the calibrated spectra were summed.

\begin{figure}
	\centering
	\includegraphics[width=6cm, angle=0]{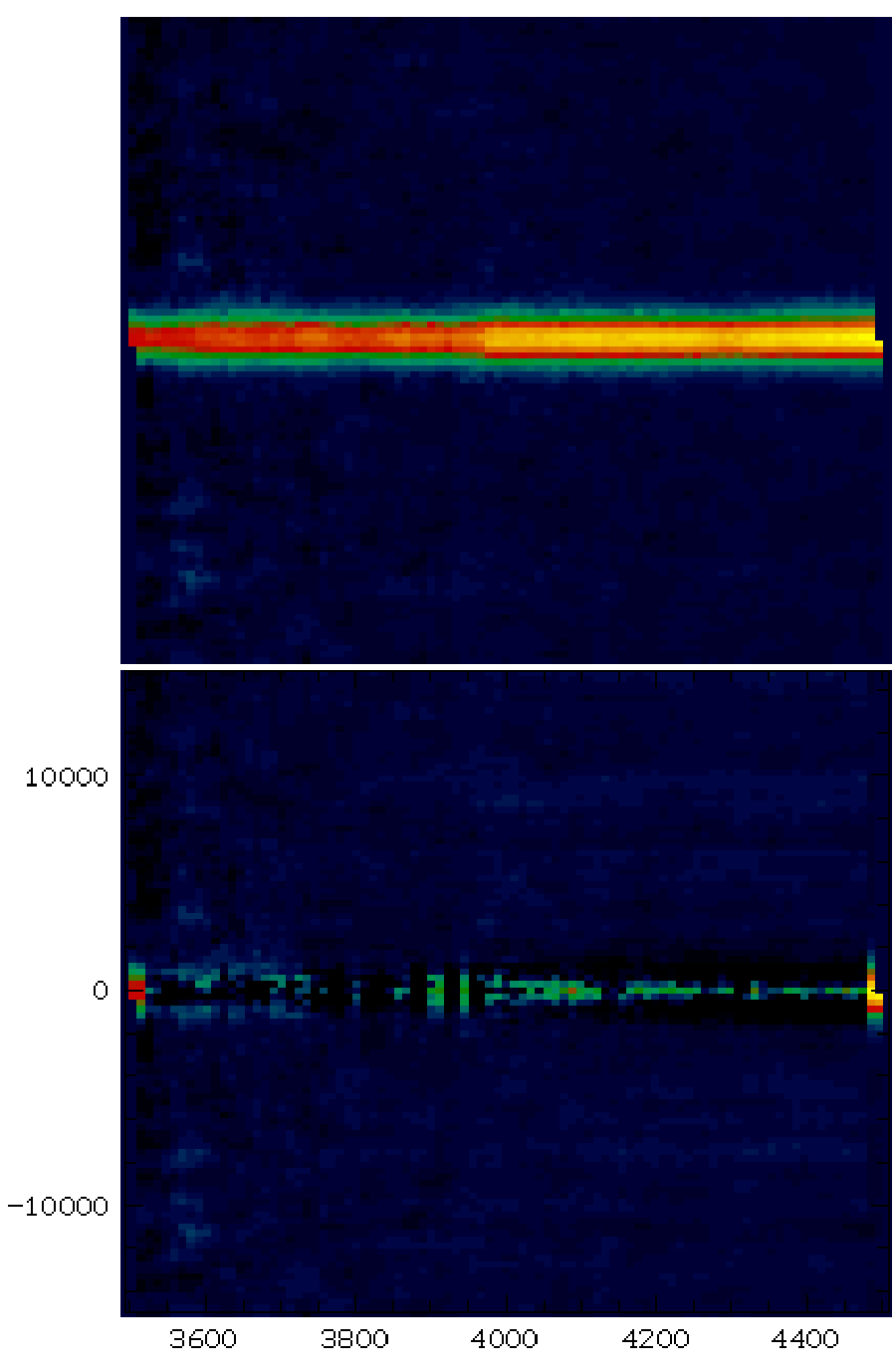}
	\caption{Combined spectrum of the comet in the region of the CN band (top) and that of the  `gas", after the subtraction of  the dust spectrum (bottom). The intensity scale is the same for both and the look up table is logarithmic. The spectral region is between  3500 and 4500 \AA~ and the spatial region is from -15000 to 15000 km of projected nucleocentric distance ($\rho$). Notice that no emission bands are seen within the SNR. The region very close to the comet photometric center is very noisy, but already few thousands km farther the SNR is much better.
}
	\label{Fig7}
\end{figure}

 A two dimensional cometary dust spectrum was obtained and subtracted from the 133P spectrum. First, the`color" (not the real dust color because the observations were not made in parallactic angle ) of the dust was found by comparing the solar analogue (SA) and comet spectra in regions out of possible gas emission. Then the dust profile along the slit was measured in the same regions. The final 2D dust spectrum is composed by the colored SA spectrum and the profile of the 133P dust. The subtraction gives the cometary gas spectrum shown in Figure~\ref{Fig7}. Notice that no emission bands are detectable in all the spectral range within the uncertainties. 

In order to have an upper limit of the CN bands intensity, the standard deviation was measured in correspondence with the CN bands in two regions at projected nucleocentric distance of 3000 km, over a width of 2000 km by 26 \AA. The spectrum has the highest SNR in those regions, since the dust contribution is already very low. Each region has a standard deviation $\sigma = 3.6 \times 10^{-11}$ erg/(cm$^{2}$ ster \AA\ s), that, considering the $\Delta \lambda$ = 26 \AA~ and the use of 2 regions, gives a total $ \sigma = 6.6 \times 10^{-10} $ erg/(cm$^{2}$ ster s). Assuming a limit of $3 \sigma$ the upper limit of the intensity is $2 \times 10^{-9}$ erg/(cm$^{2}$ ster s).  With a CN g-factor of $0.36 \times 10^{-13}$ erg/(s~ mol) (Sleicher \cite{Schleicher83}) the column density is $6.4\times10^5$ cm$^{-2}$. 



Using the vectorial model (Festou, \cite{Festou81}) the corresponding upper limit of CN production rate Q(CN) was computed. It is $ 2.8 \times 10^{21}$ mol/s, about three orders of magnitude lower than the Q(CN) determined for Jupiter family comets observed at similar heliocentric distances (A'Hearn et al. \cite{Ahearnetal95}). The upper limit of CN mass loss is $1.2 \times 10^{-4}$ kg/s. Assuming a Q(CN/H$_2$O)$\sim$0.001 (A'Hearn et al. \cite{Ahearnetal95}) gives a water production rate Q(H$_2$O) $= 2.7 \times 10^{24}$ mol/s, corresponding to a water mass loss $8 \times 10^{-2}$ kg/s.  

Excluding the contribution of the tail, Bagnulo et al. (\cite{Bagnulo2010}) give a diffuse $Af\rho$ of the dust of the
order of 0.6-0.9 cm. Assuming the empiric formula given in (Kidger \cite{Kidger04}) a rough approximation of the dust mass loss of the order of 0.7-1.6 kg/s can be derived. 
The  $Af\rho$/Q(CN) $> 2-3 \times 10^{-22}$ cm/(mol~s) is compatible with the values determined for other Jupiter family comets observed at similar heliocentric distances (A'Hearn et al. \cite{Ahearnetal95}).

\section{Discussion and conclusions}


We present the visible spectra of MBCs 133P/Elst-Pizarro and 176P/LINEAR, and of three Themis family asteroids (62) Erato, (379) Huenna and (383) Janina. 

The spectra of 133P and 176P are very similar to each other and to those of B-type asteroids. The measured spectral slopes in the visible are $S'_V$ = 0.00 $\pm 0.01$~\%/1000\AA~ and -0.03 $\pm 0.01$~\%/1000\AA~ for 133P and 176P respectively. 

The spectra of 133P and 176P are very similar to those of the three Themis family asteroids, supporting the hypothesis that 133P and 176P are Themis family members and unlikely are interlopers that originated in any of the known cometary reservoirs (the Oort Cloud and the trans-neptunian belt). 

To determine if water ice sublimation is the mechanism that produces the observed activity of MBCs, the spectra of 133P obtained with the VLT are used to check the presence of gas emission. In particular, we searched for CN emission around 0.38 $\mu$m. No CN emission is detected within the SNR of the data. An upper limit for CN production rate of  Q(CN)  $= 2.8 \times 10^{21}$ mol/s, is determined, about three orders of magnitude lower than the Q(CN) determined for Jupiter family comets observed at similar heliocentric distances (A'Hearn et al. \cite{Ahearnetal95}). That means that, if water-ice sublimation is the activation mechanism, the gas production rate is very low and/or the parent molecules of CN present in the nuclei of normal comets are much less abundant in the MBCs.

The spectra of the MBCs presented in this paper are very different from that of comet 162P/Siding-Spring and most other comet nuclei. Considering that most cometary nuclei and asteroids in cometary orbits (likely dormant comets) have spectra and colors typical of P- and D-type asteroids, while MBCs are B-type, we conclude that MBCs and comets are taxonomically different. This argues against a cometary origin of these MBCs.

On the other hand, both MBCs have a spectrum that is very similar to that of the best studied activated asteroids in the near-Earth population, (3200) Phaethon. This supports the hypothesis that  ``activated asteroids" in the NEA population are scattered MBCs (Meng et al. \cite{Mengetal04}, Licandro et al. \cite{Licandroetal07}). Notice that (3200) Phaethon's surface composition is not compatible with a cometary origin (Licandro et al. \cite{Licandroetal07}), which also supports the asteroidal nature of these MBCs. If the activity is water-ice driven, these results suggest that there are some ``activated asteroids" in the NEA and main belt population that were able to retain water ice that sublimates under certain circumstances.

Additionally, (3200) Phaethon is likely an scattered asteroid from the Pallas family (de Le\'on et al. \cite{leon2010}), while 133P and 176P are Themis family asteroids. Therefore, two large B-type asteroids, (2) Pallas and (24) Themis, are likely parent bodies of asteroids that could have retained some volatiles and present cometary-like activity. Exploring the volatile content of icy minor bodies is critical for understanding the physical conditions and the mechanisms of planetary formation, and also addresses the question of the origin of Earth's water. If the outer main belt has a large population of asteroids with ice, they could have contributed to the water on Earth. Finally, this indicates the extent and origin of volatiles in asteroids that could be used as resources for space exploration.

{\em Acknowledgements:}
This article is based on observations made with the WHT, TNG and NOT telescopes operated on the island of La Palma by the ING, FFG--INAF and NOTSA respectively,  in the Spanish ``Observatorio del Roque de los Muchachos", and with the VLT . 
JL gratefully acknowledges support from the spanish ``Ministerio de Ciencia e Innovaci\'on'' project AYA2008-06202-C03-02.
NPA acknowledges support from NASA Postdoctoral Program, administered by Oak Ridge Associated Universities through a contract with NASA.
HC gratefully acknowledges support from NASA and NSF.


\begin{thebibliography}{}
\bibitem[1995]{Ahearnetal95}
A'Hearn, M. C., Millis, R. L., Schleicher, D.,G., Osip, D.,J., Birch, P. V., 1995,
Icarus, 118, 223
\bibitem[2010]{Bagnulo2010}
Bagnulo, S., Tozzi, G. P., Boehnhardt, H., Vincent, J.-B., Muinonen, K. 2010, A\&A 514, 99.
\bibitem[1989]{Bell1989}
Bell, J. 1989, Icarus, 78, 426.
\bibitem[1997]{Boehnhardt1997}
Boehnhardt, H., Sekanina, Z., Fiedler, A., et al. 1997, Highlights Astron., 11A,
233
\bibitem[2002]{Bottkeetal02}
Bottke, W. F., Morbidelli, A., Jedicke, R., Petit, J-M., Levison, H. F., Michel, P., \& Metcalfe, T. S., 2002, Icarus, 156, 399
\bibitem[1989]{Bowel89}
Bowell, E., Hapke, B., Domingue, D., Lumme, K., Peltoniemi, J., \& Harris, A.~W.\ 1989, in  Asteroids II, Binzel, R.P., Gehrels, T., Matthews, M.S. (eds.), Univ. of Arizona Press, Tucson,  524 
\bibitem[1992]{Bowell92} 
Bowell, E., West, R., Heyer, H. et al., 1992, IAUC 5585
\bibitem[1999]{Bus99}
Bus, S. J. 1999, Ph.D. Thesis, Massachusetts Institute of Technology
\bibitem[1998]{CampSwin98} 
Campins, H., \& Swindle, T., 1998, Met. \& Plan. Sci. 33, 1201
\bibitem[1995]{Campins95} 
Campins, H., Osip, D.~J., Rieke, G.~H., \& Rieke, M.~J.\ 1995, P\&SS, 43, 733 
\bibitem[2006]{campins2006}
Campins, H., Ziffer, J., Licandro, Javier; Pinilla-Alonso, N.,  Fern\'andez, Y., de Le\'on, J., Moth\'e-Diniz, T., Binzel, R., 2006, AJ, 132, 1346.
\bibitem[2007]{campins2007}
Campins, J., Licandro, Javier; Pinilla-Alonso, N.,  H., Ziffer,  de Le\'on, J., Moth\'e-Diniz, T., Guerra, J.C., Hergenrother, C., 2007, AJ, 134, 1626.
\bibitem[2009]{Campinsetal09}
Campins, H., Kelley, M. S., Fern\'andez, Y.,  Licandro, J. \& Hargrove, K., 2009, EM\&P, 105, 159.
\bibitem[2010]{campins2010}
Campins, H., Hargrove, K., Pinilla-Alonso, N., Howell, E., Kelley, M., Licandro, J., Moth\'e-Diniz, T., Fern‡ndez, Y. \& Ziffer, J. 2010, Nature, 464, 1320
\bibitem[1982]{Carusi1982}
Carusi, A., Valsecchi, G. 1982, A\&A 115, 327.
\bibitem[1996]{Chamberetal96}
Chamberlin, A.~B., McFadden, L.-A., Schulz, R., Schleicher, D.~G., \& Bus, S.~J.\ 1996, Icarus, 119, 173 
\bibitem[2010]{Clark2010}
Clark, B., Ziffer, J., Nesvorny, D. et al. 2010, JGRE 115, E06005, 22.
\bibitem[1950]{Cunningham50}
Cunningham, L.~E.\ 1950, \iaucirc, 1250, 3 
\bibitem[2003]{Delboetal03}
Delb\'o, M., Harris, A.W., Binzel, R.P., Pravec, P., Davies, J.K., 2003. Icarus, 166, 116.
\bibitem[2004]{Delbo04}
Delb\'o, M., 2004. Ph.D. thesis, Freie Universit{\"a}t, Berlin.
\bibitem[2007]{Delboetal07}
Delb\'o, M., dell'Oro, A., Harris, A.W., Mottola, S., \& M{\"u}ller, M. 2007, Icarus 190,
236
\bibitem[2010]{leon2010}
de Le\'on, J., Campins, H., Tsiganis, K., Morbidelli, A., Licandro, J. 2010, A\&A 513, 26.
\bibitem[2011]{deLeon2011}
de Le\'on, J., Licandro, J., Serra-Ricart, M., Pinilla-Alonso, N., Campins, H.,  2011, A\&A 517, 23.
\bibitem[2006]{Drake06}
Drake M. J. \& Campins H., 2006, In Asteroids, Comets and Meteorites, Cambridge Univ. Press,
p. 381.
\bibitem[1996]{Elst96} 
Elst, E., Pizarro, O., Pollas, C., Ticha, J., Tichy, M., Moravec, Z., Outt, W., Marsden, B., 1996, IAUC 6456.

\bibitem[2008)]{Emery08} 
Emery, J.~P., Lim, L.~F., Marchis, F., \& Cruikshank, D.~P.\ 2008, LPI Contributions, 1405, 8345 
\bibitem[1997]{Fernandezetal97}
Fern\'andez, Y., McFadden, L., Lisse, C., Helin, E., \& Chamberlin A., 1997, Icarus, 128, 114.
\bibitem[2002]{Fernandez2002} 
Fern\'andez, J., Gallardo, T., and Brunini, A. 2002, Icarus 159, 358-368.
\bibitem[2005]{Fernandez05} 
Fern{\'a}ndez, Y., Jewitt, D., \& Sheppard, 2005, \aj, 130, 308
\bibitem[2008]{Fernandezetal08} 
Fern{\'a}ndez, Y.~R., et al.\ 2008, LPI Contributions, 1405, 8307 
\bibitem[1981]{Festou81}
Festou, M., C., 1981, A\&A, 95, 69
\bibitem[1999]{Florczak1999}
Florczak, M., Lazzaro, D., Moth\'e-Diniz, T., Angeli, C.,
\& Betzler, A., 1999, A\&ASS, 134, 463
\bibitem[1989]{Gusta89}
Gustafson, B., 1989, A\&A 225, 533
\bibitem[2009]{Groussinetal09} 
Groussin, O., et al.\  2009, Icarus, 199, 568 
\bibitem[2009]{Haghighipour}
Haghighipour, N. 2008. M\&PS, 44, 1863.

\bibitem[1998]{Harris1998}
Harris. A. W., 1998, Icarus, 131
\bibitem[2006]{Hsieh04}
Hsieh, H., Jewitt, D.,  Fern\'andez, Y., 2004, AJ 127, 299.

\bibitem[2006]{Hsieh06}
Hsieh, H., \& Jewitt, D., 2006, Science 312, 561.
\bibitem[2009]{Hsieh09}
Hsieh, H.,  Jewitt, D., \& Fern\'andez, Y., 2009, ApJ, 694, 111. 
\bibitem[1010]{Hsieh10}
Hsieh, H., Jewitt, D., Lacerda, P., Lowry, S., \& Snodgrass, C., 2010, MNRAS, 403, 363.
\bibitem[2002]{Ivezic2002}
Ivezic, Z. et al. 2002, AJ, 124, 2943.
\bibitem[1986]{Jakosky}
Jakosky, B.M., 1986, Icarus, 66, 117.
\bibitem[2002]{jewitt2002} 
{Jewitt, D.} 2002, \textit{A.J}, 123, 1039-1049.
\bibitem[2004]{Kidger04}
Kidger, M., R., 2004, A\&A, 420, 389
\bibitem[1984]{Keihm}
Keihm, S.J., 1984, Icarus 60, 568.
\bibitem[1992]{Kosai92}
Kosai, H. 1992, Celest. Mech. Dyn. Astron., 54, 237.
\bibitem[2005]{Kraemer05} 
Kraemer, K.~E., Lisse, C.~M., Price, S.~D., Mizuno, D., Walker, R.~G., Farnham, T.~L., Makinen, T.\ 2005, \aj, 130, 2363 
\bibitem[1982]{Kresak82}
Kresak, L. 1982, Bull. Astron. Inst. Czech., 33, 104.
\bibitem[1989]{Kuhrt} 
K{\"u}hrt, E., Giese, B., 1989, Icarus 81, 102
\bibitem[2009]{Lamy2009}
Lamy, P. \& Toth, I. 2009, Icarus, 201, 674.
\bibitem[1992]{landolt}
Landolt A., 1992,
AJ, 104, 340-491.
\bibitem[2006]{Levisonetal2006} 
Levison, H.~F., Terrell, D., Wiegert, P.~A., Dones, L., \& Duncan, M.~J.\ 2006, Icarus, 182, 161 
\bibitem[2009]{levison2009}  
Levison, H., Bottke, W., Gounelle, M., Morbidelli, A., Nesvorny, D. \& Tsiganis, K. 2009, Nature, 460, 364.


\bibitem[2002]{licandro2002}  
{Licandro, J., Campins, H., Hergenrother, C., Lara, L. M.} 2002, \textit{A\&A}, 398, L45-L50.
\bibitem[2007]{Licandroetal07}
Licandro, J., Campins, H., Moth\'e-Diniz, T., Pinilla-Alonso, N., de Le\'on, J., 2007, A\&A, 461, 751.
\bibitem[2008]{licandro08}
Licandro, J., Alvarez-Candal, A., de Le\'on, J., et al. 2008, \aap, 481, 861
\bibitem[2005]{Makovoz05} 
Makovoz, D., \& Khan, I.\ 2005, Astronomical Data Analysis Software and Systems XIV, 347, 81 
\bibitem[1995]{Marzari1995}
Marzari, F., Farinella, P., Vanzani, V. 1995, A\&A 299, 267.
\bibitem[2004]{Mengetal04}
Meng, H., Zhu, J., Gong, X. et al.,  2004, Icarus 169, 385
\bibitem[2005]{Moth2005}
Mothe-Diniz, T.,  Roig, F. \& Carvano, J. M., 2005, Icarus, 174, 54
\bibitem[2007]{Mottl} 
Mottl, M., Glazer, B., Kaiser, R., \& Meech, K.,  Chemie der Erde Geochemistry, 67, 253
\bibitem[2007]{MullerPhd} 
Mueller, M., 2007. Surface Properties of Asteroids from Mid-Infrared 
Observations and Thermophysical Modeling. Digitale Dissertation, 
Freie Universitaet Berlin. Available from: 
http://www.diss.fu-berlin.de/2007/471/indexe.html 
\bibitem[2007]{Muller07} 
M{\"u}ller, T.~G., \& Lagerros, J.~S.~V.\ 1998, \aap, 338, 340 
\bibitem[2008]{Nesvorny2007}
Nesvorny, D., Bottke, W., Vokrouhlicky, D., Sykes, M., Lien, D., Stansberry, J., 2008, ApJ 679, L143.
\bibitem[2006]{Ohtskuaetal06}
Ohtsuka, K., Sekiguchi, T., \& Kinoshita, D. et al., 2006,
A\&A 450, 250
\bibitem[1995]{Osip95} 
Osip, D., Campins, H., \& Schleicher, D.~G.\ 1995, Icarus, 114, 423 
\bibitem[2007]{Reach07} 
Reach, W.~T., Kelley, M.~S., \& Sykes, M.~V.\ 2007, Icarus, 191, 298 
\bibitem[2010]{rivkin2010}  
Rivkin, A.S. \& Emery, J. 2010, Nature, 464, 1322.
\bibitem[2011]{Rousselotetal11}
Rousselot, P., Dumas, C., \& Merlin, F. 2011, Icarus, 211, 553.
\bibitem[1983]{Schleicher83} 
Schleicher, D. G., 1983, Ph.D. Thesis Maryland Univ., College Park.
\bibitem[2008]{Schorghofer} 
Schorghofer, N., 2008, ApJ, 682, 697.
\bibitem[2008]{Snodgrass08} {Snodgrass, C., Lowry, S.~C., \& Fitzsimmons, A.} 2008, \textit{MNRAS}, 385, 737 
\bibitem[1990]{Spencer}
Spencer, J.R., 1990, Icarus 83, 27
\bibitem[1999]{Tanga1999}
Tanga, P., Cellino, A, Michel, O., Zappala, V., Paolicchi, P., dell'Oro, A. 1999, Icarus, 141, 65.
\bibitem[1989]{Tholen1989}
Tholen, D. J. 1989, in  Asteroids II, Binzel, R.P., Gehrels, T., Matthews, M.S. (eds.), Univ. of Arizona Press, Tucson,  298
\bibitem[2006]{Toth06}
Toth, I., A\&A, 2006, 446, 333.

\bibitem[1983]{Whiple93}
Whipple, F., 1983, IAUC 3881
\bibitem[1993]{WillWu93}
Williams, I.P. \& Wu, Z., D., 1993, MNRAS 264, 659
\bibitem[1990]{Zappala1990}
Zappal\`a, V., Cellino, A, Farinella, P., Knezevic, Z. 1990, AJ, 100, 2030. 
\end{thebibliography}
\end{document}